\documentclass[journal=jacsat,manuscript=article]{achemso}

\SectionNumbersOn

\usepackage{amsmath}
\usepackage{amsfonts}
\usepackage{graphics}
\usepackage{physics}
\usepackage{xr}
\usepackage{breqn}
\usepackage{textcomp}
\usepackage{xcolor}
\usepackage{pdfpages}

\title{Scattering-matrix approach for a quantitative evaluation of the topological protection in valley photonic crystals}

\author{Ga\"etan L\'ev\^eque}
\affiliation{Institut d'Electronique, de Micro-\'electronique et de Nanotechnologie (IEMN, CNRS-8520), Cit\'e Scientifique, Avenue Poincar\'e, 59652 Villeneuve d'Ascq, France}
\email{gaetan.leveque@univ-lille.fr}
\author{Yan Pennec}
\affiliation{Institut d'Electronique, de Micro-\'electronique et de Nanotechnologie (IEMN, CNRS-8520), Cit\'e Scientifique, Avenue Poincar\'e, 59652 Villeneuve d'Ascq, France}
\author{Pascal Szriftgiser}
\affiliation{Université de Lille, CNRS, UMR 8523-PhLAM-Physique des Lasers Atomes et Molécules, F-59000 Lille, France}
\author{Alberto Amo}
\affiliation{Université de Lille, CNRS, UMR 8523-PhLAM-Physique des Lasers Atomes et Molécules, F-59000 Lille, France}
\author{Alejandro Mart\'inez}
\affiliation{Nanophotonics Technology Center, Universitat Politècnica de València, Camino de Vera s/n, 46022 Valencia, Spain}

\begin{document}
	
\begin{abstract}
The realization of photonic crystal waveguides with topological protection enables robust light propagation against defect-induced scattering. It should allow the design of very compact devices by exploiting guiding through sharp bends with low losses and back-reflection. In this work, we use valley-topological triangular resonators coupled to an input waveguide to evaluate the quality of the topological protection. To that purpose, we first analyze via numerical simulations the existence of backward scattering at cavity corners or transmission with pseudo-spin conversion at the splitter between the input waveguide and the cavity. We evidence that a breakdown of topological protection takes place, in particular at sharp corners, which results in transmission minima and split-resonances, otherwise non-existent. In order to evaluate the small coupling coefficients associated to this breakdown, a phenomenological model based on an \textcolor{black}{exact} parameterization of scattering matrices at splitters and corners of the resonators is then introduced. By comparison with the numerical simulations, we are able to quantify the loss of topological protection at sharp bends and splitters. Finally, \textcolor{black}{we use the obtained set of phenomenological parameters to compare the predictions of the phenomenological model with full numerical simulations for fractal-inspired cavities based on the Sierpi\'nski triangle construction. We show that the agreement is overall good, but shows more differences for the cavity composed of the smallest triangles.} Our results suggest that even in a system exempt of geometrical and structural defects, topological protection is not complete at corners, sharp bends and splitters. \textcolor{black}{However, simpler but predictive calculations can be realized with a phenomenological approach, allowing simulations of very large devices beyond the reach of standard simulation methods,}  which is crucial to design photonic devices which gather compactness and low losses through topological conduction of electromagnetic waves.
\end{abstract}

Topological photonics has recently become a disruptive paradigm enabling exotic ways to manipulate light propagation \cite{khanikaev_photonic_2013,ozawa_topological_2019,segev_topological_2020,iwamoto_recent_2021,price_roadmap_2022}. Amongst the different platforms to implement photonic structures relying on topological effects, two-dimensional (2D) high-index photonic crystal slabs display interesting features such as lossless propagation and large bandwidth, compatible with standard microfabrication tools \cite{barik_topological_2018,shalaev_robust_2019,he_silicon--insulator_2019,parappurath_direct_2020,arora_breakdown_2022,barczyk_interplay_2022,kumar_phototunable_2022}. An interesting proposal to build a topologically-protected waveguide in a 2D photonic crystal was presented in Wu et al\cite{wu_scheme_2015}. Essentially, the idea is to design the unit cell of a honeycomb lattice so that it shows a Dirac cone at the $\Gamma$ point at a given frequency. Then either by shrinking or expanding the motif inside unit cells, a topological band-gap arises. The interface between two semi-infinite shrunken and expanded lattices supports topologically protected modes showing a certain pseudo-spins for a given propagation directions \cite{wu_scheme_2015}. Remarkably, when this approach is applied to 2D photonic crystal slabs, the guided modes are always over the light line, meaning that they are always radiative, a property that has been used to identify the pseudo-spin of the guided modes via far-field measurements \cite{parappurath_direct_2020,arora_breakdown_2022}. 

The realization of large-scale photonic integrated circuits requires, however, waveguides that do not radiate. In photonic crystal slabs, this means that the guided modes should be below the light line to ensure perfect confinement by total internal reflection. The realization of topological waveguides supporting fully guided modes would require thus a honeycomb lattice showing - when undeformed - a Dirac point at symmetry points different to $\Gamma$ in the first Brillouin zone. 
In contrast to the shrunken-expanded configuration, which mimics the spin Hall effect for photons, it has been proposed \cite{ma_all-si_2016,noh_observation_2018} and experimentally realized \cite{wu_direct_2017,shalaev_robust_2019,he_silicon--insulator_2019} photonic analogs of the valley Hall effect. Since experimental works use standard silicon technology, valley Hall photonic waveguides show a huge potential to become key elements in silicon photonics. One of the great advantage over light waveguiding along line defects in trivial photonic crystals is the ability, thanks to the topological protection, to conduct light even along sharp corners, with angles as small as 60$^o$ \cite{shalaev_robust_2019,he_silicon--insulator_2019,ma_topological_2019,arora_direct_2021}. This would allow for increasing the compacity and decreasing the footprint of future devices for information and communication technologies. However, and unlike in other topological photonic systems in which the time-reversal symmetry is broken (for instance by applying an external magnetic field \cite{wang_observation_2009,bahari_nonreciprocal_2017}), in photonic crystals the band structure is symmetric with respect to the wave vector along the propagation direction. This means that for any topologically-protected guided mode having a certain \textcolor{black}{helicity relying its} pseudo-spin and wave vector, there will be an identical state with opposite \textcolor{black}{helicity}. Albeit the topology of the system provides certain robustness to the propagation \cite{shalaev_robust_2019,he_silicon--insulator_2019}, backward scattering is not prohibited by nature. \textcolor{black}{Few works concern the assessment of the robustness of the topological protection\cite{arregui_quantifying_2021,rosiek_observation_2023}, and they essentially concern the back-reflection induced by inhomogeneities inside the photonic crystal itself, which as a result introduces propagation losses distributed over the whole length of the topological edge. To our knowledge, a quantitative evaluation of the pseudo-spin conversion in \textit{perfect} topological crystals} in presence of corners or other guiding elements like splitters \textcolor{black}{is still lacking, and} is a mandatory step to assess the applicability of such waveguides to more complex photonic systems, where those processes can be highly detrimental. This is particularly the case of ring-like cavities in which the finesse is highly sensitive to small losses due to the continuous recirculation of light.

In this work we analyze via numerical simulations the properties of valley topological edge-modes built in 2D photonic crystals. In particular, we focus on the loss of topological protection at sharp corners and splitters of triangular resonators coupled to a linear waveguide\cite{siroki_topological_2017,barik_chiral_2020,jalali_mehrabad_chiral_2020,monika_devi_topological_2021}, either from a corner or the middle of an edge. In such systems, in absence of absorption, no resonant features are expected in the transmission spectra if perfect topological protection is realized: any deviation from a flat transmission band can in principle be traced back to a breakdown of topological protection somewhere along the path of light. We show that, in triangular resonators with different coupling conditions, transmission spectra present minima and split-resonances due to the coupling between counter-propagating waves, as in the case of other non-topological traveling-wave resonators \cite{zhang_resonance-splitting_2008}. In order to elucidate and quantify the origin of the phenomenon, we then introduce a phenomenological model relying on the description of corners and splitters by scattering matrices \textcolor{black}{whose exact expressions are derived. The numerical evaluation of the eight real parameters describing those matrices allow reproducing very precisely} the simulations \textcolor{black}{and assessing the small, frequency-dependent, coupling coefficients corresponding to pseudo-spin conversions at corners and splitters}. Our results evidence that, even in the case of propagation of light in topological circuits \textcolor{black}{free from geometrical and structural imperfections}, reflection at sharp corners dominates the \textcolor{black}{overall} shape of transmission spectra, \textcolor{black}{whose finer details are attributed to weaker ruptures of topological protection at the splitter}. We conclude our study by \textcolor{black}{demonstrating that our phenomenological model allows predictive and faster numerical simulations of complex circuit, taking the example of fractal-inspired resonators based on a Sierpi\'nski triangle construction.}
\begin{figure*}[!h]
	\includegraphics[width=15cm]{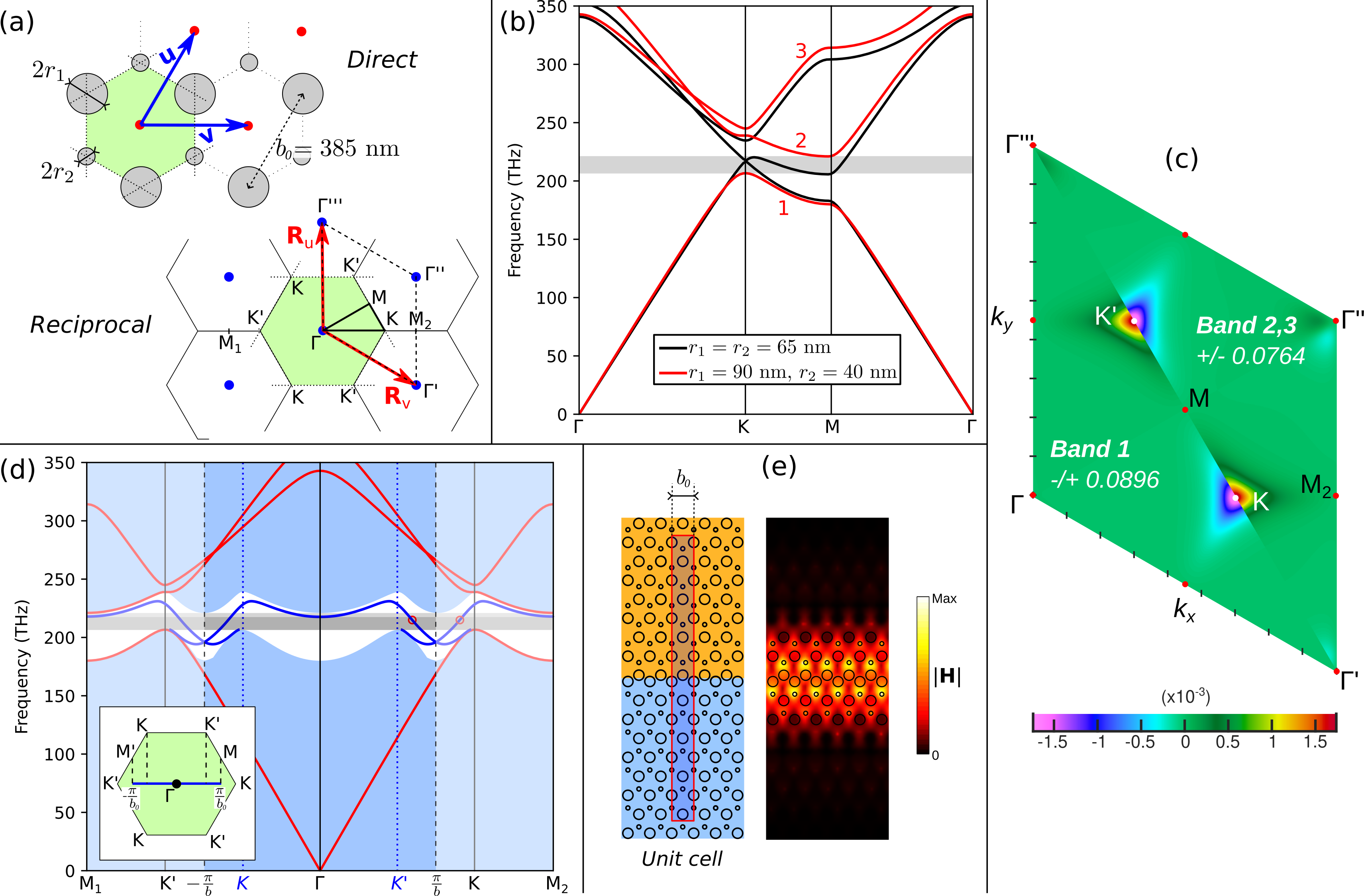}
	\caption{(a) Geometry of the TPC and representation of the reciprocal space and associated high-symmetry points. (b) Dispersion diagram of the TPC for equal (black) or different (red) radii $r_1$ and $r_2$ of the two holes inside each primitive cell. (c) Berry curvature and valley Chern numbers simulated for the disymmetric TPC ($r_1=180$ nm and $r_2=80$ nm). (d) Dispersion curves (solid blue lines) of the edge modes propagating along the bearded interface in between two semi-infinite mirror-symmetric TPCs, parallel to the $\Gamma K$ direction (the light blue background indicates the projected bulk modes). The solid red lines show the dispersion curves of the infinite TPC. The inset compares the FBZ of the interface (thick blue line with length $2\pi/b_0$) and the FBZ of the infinite TPC. (e) Typical unit cell used in the simulation (left panel) and distribution of magnetic field amplitude of the edge mode (right panel).}
	\label{fig:disp2d}
\end{figure*}

\section{Numerical approach}
Our topological photonic crystal (TPC), see Fig.~\ref{fig:disp2d}(a), is based on the well-known hexagonal-lattice geometry made of circular holes with radii $r_1$ and $r_2$, investigated for example in He et al\cite{he_silicon--insulator_2019} and other works \cite{barik_chiral_2020,jalali_mehrabad_chiral_2020}. The study is restricted to TE (in-plane) polarization. For our bi-dimensional system, we chose a lattice constant $b_0=385$ nm, an average hole radius $r_0=130$ nm, and a refractive index $n=2.7$. Those values allow matching the band-gap of the silicon membrane described in  He et al\cite{he_silicon--insulator_2019}.  Unless specified, all numerical simulations have been performed using the finite-elements-method (FEM) software Comsol Multiphysics. When $r_1=r_2=r_0$, the band diagram of the resulting honeycomb lattice presents Dirac cones at the six $K$ points at the edge of the first Brillouin zone (FBZ), close to $f_0=216$ THz (Fig.~\ref{fig:disp2d}(b), black line). A band-gap is then opened around $f_0$ for $r_1\ne r_2$ (Fig.~\ref{fig:disp2d}(b), red line), where 6-fold rotational symmetry of the lattice point group is lowered to 3-fold due to the breaking of inversion symmetry. For $r_1=180$ nm and $r_2=80$ nm, the band-gap corresponds to the frequency window [205.7 THz, 220.7 THz], indicated by the gray area.

The bands surrounding this bandgap present a non trivial local topology as expected from the valley Hall effect. This can be readily seen by computing the Berry curvature of those bands using a planewave expansion method \cite{blanco_de_paz_tutorial_2020} (see details in section 1 of the supplementary informations file). Figure \ref{fig:disp2d}(c) shows the calculated Berry curvature for the lowest band (lower right corner of the figure) and for the ensemble of bands 2 and 3 (upper right corner), which are touching. In both cases, the Berry curvature is concentrated at the $K$ and $K'$ points. For a given band, it has opposite sign at the $K$ and $K'$, as expected for a time reversal symmetric system, and at a given $K/K'$ point, each set of bands presents Berry curvatures of opposite sign. This configuration of opposite signs at opposite $K/K'$ points and different bands is at the origin of the interface topological modes when two mirror symmetric photonic crystals are pasted together. This is confirmed by the non-zero values of valley Chern numbers, calculated by integrating the Berry curvature around the $K$ and $K'$ points: we obtain $\mp 0.090$ for the first band and $\pm 0.076$ for the second and third bands. The computed Chern numbers are low compared to the usually expected values of $\pm1/2$. Actually, as mentioned by several authors\cite{he_silicon--insulator_2019,wong_gapless_2020}, the Chern number is a half integer in the limit of weak perturbations, which corresponds here to small dissymmetries of the holes radii. In order to open an appreciable band-gap, $r_1$ must be significantly different of $r_2$, which leads to an overlap of the Berry curvatures with opposite signs in each half of the 1BZ, finally resulting in lower Chern numbers\cite{he_silicon--insulator_2019}.

We show on Fig.~\ref{fig:disp2d}(d) the dispersion relation of the topological edge mode propagating along a bearded $\Gamma K$ edge, in between two semi-infinite TPCs with glide mirror symmetry (in orange and blue on Fig.~\ref{fig:disp2d}(e)). The system being now uni-dimensional with period $b_0$, its FBZ is a segment with length $2\pi/b_0$ along the $\Gamma K$ direction (see inset), fully included inside the FBZ of the TPC. The breakdown of periodicity along the direction perpendicular to the edge direction induces a projection of the bulk modes of the infinite TPC onto the linear FBZ, which correspond to the light blue background, and of $K$ and $K'$ points onto the points indicated by the blue dotted lines. The shape of a typical unit cell used in the numerical simulation is plotted on Fig.~\ref{fig:disp2d}(e), left panel, together with the distribution of the magnetic field amplitude for a frequency of 215.6 THz : the field is concentrated at the interface between the two crystals with a penetration length of about 1.5 unit cells into the bulk. The frequency of the topological mode has a local minimum at the $\Gamma$ point, with a frequency of about 217.1 THz, lower than the bottom of the bulk band-gap. For that reason, the effective band-gap for the topological edge mode is [205.7 THz, 217.1 THz], underlined in dark gray on Fig.~\ref{fig:disp2d}(d). From this simulation, we can extract the evolution of the wavevector or, equivalently, the effective index of the topological edge mode with frequency.

\begin{figure}[!h]
	\includegraphics[width=7cm]{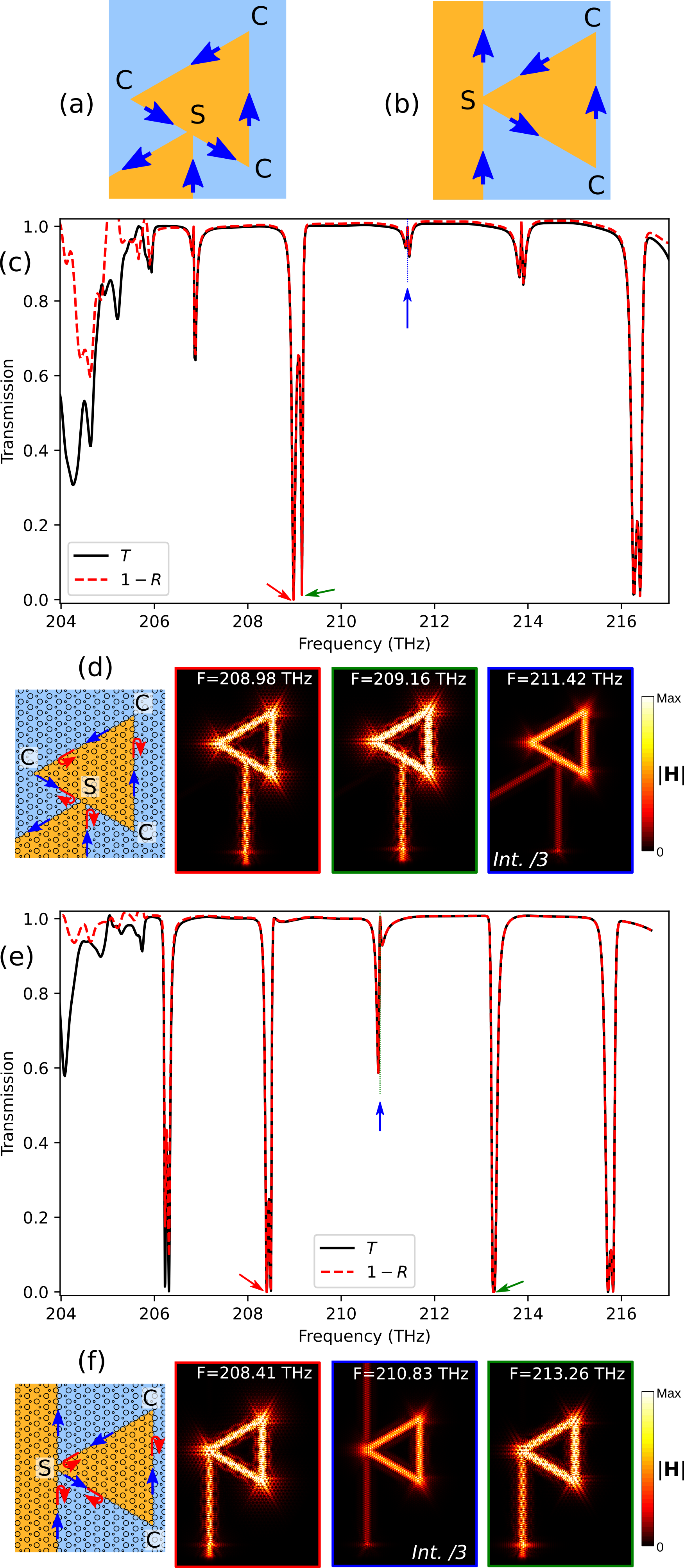}\\
	\caption{(a) and (b), respective representations of an edge- or corner-addressed topological triangular cavity. (c) Comparison between the transmission (solid black line) and reflection (dashed red line) spectra obtained from numerical simulations for the edge-addressed cavity. (d) Distribution of the magnetic field amplitude for frequencies indicated by a colored arrow on (c). For $F=211.42$ THz, the maximum value is three times larger as for the two first frequencies. (e) and (f): same as (c) and (d) but for the corner-addressed cavity.}
	\label{fig:meta}
\end{figure}

\section{Triangular resonators}
To investigate the robustness of the topological protection, we now characterize the properties of triangular edge mode resonators coupled to waveguides built on $\Gamma K$ bearded edges. Such cavities can be coupled to a waveguide in two different ways, either from the edge, Fig.~\ref{fig:meta}(a), or from the corner,  Fig.~\ref{fig:meta}(b). In both situations, the injection of the topological mode inside the resonator is realized through a splitter with four branches and labeled by $S$, while the corners of a triangular cavity will be noted $C$. As shown on Fig.~\ref{fig:meta}(a) and (b), if topological protection is perfect in the system (which means at $S$ and $C$ points), all topological modes propagate along each edge with the same helicity, and then along the same direction. As a consequence, the system cannot reflect waves in the excitation guide, which means that the reflection coefficient $R$ in power is zero, and through energy conservation the transmission coefficient $T$ in power is unity. Even if resonances in amplitude can occur inside the triangular cavity, they cannot have a signature in the transmission or the reflection spectra.

The numerical simulations (see section 2 in the supplementary informations file) of triangular cavities with edge length $L\approx 28b_0$ are presented on Fig.~\ref{fig:meta}(c) to (f). Notice that the TPC containing the resonators and the coupling guide has been surrounded by Perfectly Matched Layers (PML's), whose role is to absorb the field along the outer edge of the simulation domain to simulate an infinite system. The spectra show, in contrast to the previous analysis, narrow transmission dips regularly separated in frequency, both for edge- and corner- addressed resonators, however with different profiles and frequencies. The simulation domain has been taken large enough to minimize the coupling of evanescent fields emanating from the structures (for example corners) with PMLs: we can then verify numerically that $T\approx 1-R$, as black ($T$) and dashed-red ($1-R$) curves overlap in the band-gap. In order to fully explain the transmission and reflection spectra, we need to suppose the existence of losses of topological protection in the system. For that reason, the topological edge mode can travel along the directions corresponding to the same helicity as for the excitation mode (blue arrows on Fig.~\ref{fig:meta}(d) and (f)) or the opposite helicity indicated by the red arrows. This breakdown  of topological protection can originate from the splitter or the triangle corners.

Both spectra present split resonances with low transmission ($T\approx 0$, two and four respectively for the edge- and corner-addressed resonators inside the effective gap) and profiles characterized by a quasi-unity transmission ($T\approx 1$, three and one respectively) in between two transmission minima above 0.5 in power, called anti-resonances below. Split resonances can be related to mode splitting arising from back-scattering in standard ring resonators \cite{zhang_resonance-splitting_2008,zhu_-chip_2010}. Typical distributions of the magnetic field amplitude are presented on Fig.~\ref{fig:meta}(d) and (f) for split resonances and anti-resonances. For both transmission minima of the split resonance at $F\approx 209$ THz (edge-addressed cavity), a clear interference pattern is obtained inside the cavity and in the excitation guide. The difference between both distributions is visible along the bisector plane crossing each corner: along those planes, the magnetic field is maximum or minimum respectively for the lowest and highest frequencies. The frequency difference is about 160 GHz, and the full width at half maximum (FWHM) is 100 GHz and 50 GHz respectively, corresponding to quality factors of $Q=2090$ and 4180. At the anti-resonance ($F\approx 211$ THz, blue arrow), the transmission is close to unity and no interference pattern is observed accordingly in the coupling waveguide, but a small intensity modulation is visible along the triangular cavity. Notice that the colorscale is the same for all distributions, except for the anti-resonances where the maximum value of the magnetic field amplitude is three times higher. Similar observations can be made on the field distributions of the corner-addressed cavity, see Fig.~\ref{fig:meta}(f): the shape of the input waveguide does not modify the field distribution at the split- or anti- resonances, despite of their frequency shift as compared to the corner-addressed cavity.

If the occurrence of transmission split resonances and anti-resonances is a signature of a loss in topological protection, it is difficult from the numerical simulation to quantify this breakdown  and find its origin. To this purpose, we propose in the next section a phenomenological approach, where both the splitter and triangle corners are described by a scattering matrix, allowing a simplified description of the systems.

\section{Phenomenological model}
\begin{figure}[h]
	\includegraphics[width=5.5cm]{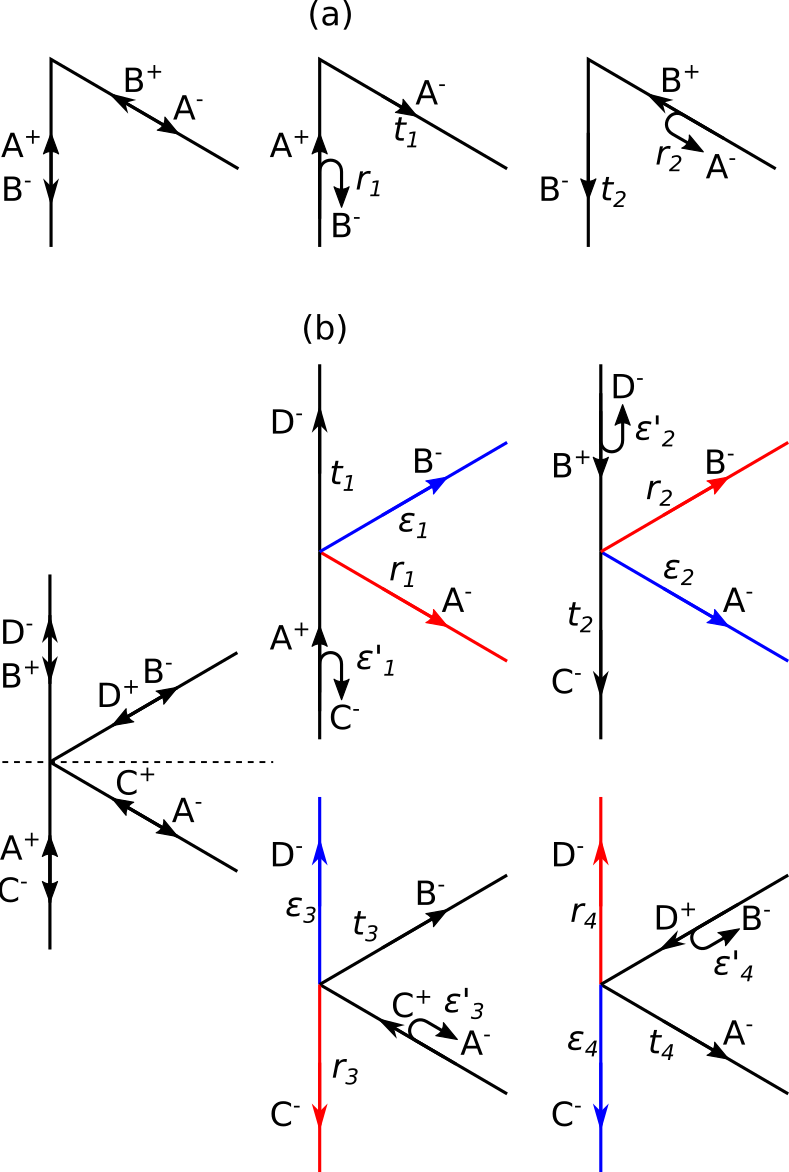}
	\caption{Definition of the coupling coefficients for the corner, (a), and the splitter, (b), as employed in the scattering matrix semi-analytical model.}	\label{fig:system}
\end{figure}

Our phenomenological model of the triangular resonators relies on the description of the splitter $S$ and corners $C$ by a scattering matrix, which expresses a linear relation between outcoming waves with amplitudes $A^{-}$, $B^{-}$, $C^{-}$ ... and incoming waves with amplitudes $A^{+}$, $B^{+}$, $C^{+}$ ... (see Fig.~\ref{fig:system}). The matrices elements can be partly extracted from numerical simulations. The topological nature of those modes implies that several of those elements are expected to be zero or much smaller than unity. The details of the calculations are given in the section 3 of the supplementary informations file, but we outline the main results below.

A simple corner, in the most general situation, behaves like a coupler between the two incident edge modes described by the complex vector $\mathbf{X}^+=\qty[A^+,B^+]^T$ with the corresponding transmitted and back-scattered modes, $\mathbf{X}^-=\qty[A^-,B^-]^T$, see Fig.~\ref{fig:system}(a). The shape of the scattering matrix, defined by the relation $\mathbf{X}^-=\mathbf{M}_C\mathbf{X}^+$, is constrained by energy conservation, which implies that $\mathbf{M}_C$ is unitary, and time-reversal symmetry, which, combined with unitary, implies that $t_1=t_2$. Finally, $\mathbf{M}_C$ have the following general form:
\begin{align*}
\mathbf{M}_C =\begin{bmatrix}
t_1 & r_1 \\ r_2 & t_2
\end{bmatrix}=e^{i\tau}
\begin{bmatrix}
\cos\sigma & i\sin\sigma \\ i\sin\sigma & \cos\sigma
\end{bmatrix}
\end{align*}
The phase $\tau$ and angle $\sigma$ can take arbitrary values. For convenience, we note in the following $a=t_1=t_2$ and $b=r_1=r_2$.

Concerning the splitter, four outputs, $\mathbf{X}^+=\qty[A^+,B^+,C^+,D^+]^T$, are now connected via the scattering matrix $\mathbf{M}_S$ to four inputs, $\mathbf{X}^-=\qty[A^-,B^-,C^-,D^-]^T$, see Fig.~\ref{fig:system}(b), with a priori 16 complex coefficients defined on the figures. However, the symmetry of the system implies that $\alpha_1=\alpha_2$ and  $\alpha_3=\alpha_4$, where $\alpha=r,\,t,\,\epsilon,\,\epsilon'$, and we will note $t_1=t_2=t$, $t_3=t_4=t'$, $\epsilon'_1=\epsilon'_2=\epsilon'$, $\epsilon'_3=\epsilon'_4=\epsilon''$. Energy conservation implies the unitarity of $\mathbf{M}_S$, and time reversal symmetry allows showing that all the $r_i$ and $\epsilon_i$ coefficients are equal. \textcolor{black}{Finally, additional symmetry considerations on the geometry of the splitter lead to the following parameterization of $\mathbf{M}_S$:}
\textcolor{black}{
\begin{multline} \label{eq:mk}
\mathbf{M}_S=
\begin{bmatrix}
r & \epsilon & \epsilon'' & t' \\
\epsilon  & r & t' & \epsilon'' \\
\epsilon'  & t & r & \epsilon \\
t & \epsilon' & \epsilon & r
\end{bmatrix}=\frac{e^{i\alpha}}{2}\times \\
\begin{bmatrix}
\qty(c_\phi\, e^{i\rho}+c_{\phi'}\, e^{-i\rho}) & \qty(c_\phi\, e^{i\rho}-c_{\phi'}\, e^{-i\rho})  & i\qty(s_\phi\, e^{-i\delta}-s_{\phi'}\, e^{i\delta})e^{-i\beta} & i\qty(s_\phi\, e^{-i\delta}+s_{\phi'}\, e^{i\delta})e^{-i\beta} \\
\qty(c_\phi\, e^{i\rho}-c_{\phi'}\, e^{-i\rho})  & \qty(c_\phi\, e^{i\rho}+c_{\phi'}\, e^{-i\rho})  & i\qty(s_\phi\, e^{-i\delta}+s_{\phi'}\, e^{i\delta})e^{-i\beta} & i\qty(s_\phi\, e^{-i\delta}-s_{\phi'}\, e^{i\delta})e^{-i\beta} \\
i\qty(s_\phi\, e^{-i\gamma}-s_{\phi'}\, e^{i\gamma})e^{i\beta} & i\qty(s_\phi\, e^{-i\gamma}+s_{\phi'}\, e^{i\gamma})e^{i\beta} & \qty(c_\phi\, e^{i\rho}+c_{\phi'}\, e^{-i\rho}) &\qty(c_\phi\, e^{i\rho}-c_{\phi'}\, e^{-i\rho})  \\
i\qty(s_\phi\, e^{-i\gamma}+s_{\phi'}\, e^{i\gamma})e^{i\beta} & i\qty(s_\phi\, e^{-i\gamma}-s_{\phi'}\, e^{i\gamma})e^{i\beta} & \qty(c_\phi\, e^{i\rho}-c_{\phi'}\, e^{-i\rho})  & \qty(c_\phi\, e^{i\rho}+c_{\phi'}\, e^{-i\rho})
\end{bmatrix}
\end{multline}
}
with the additional constraint $\gamma+\delta+2\rho=0$\textcolor{black}{, and the definitions $c_x=\cos x$ and $s_x=\sin x$}. The $\mathbf{M}_S$ matrix is then parameterized by \textcolor{black}{six} free parameters \textcolor{black}{: the two angles $\phi$ and $\phi'$, and the four phases $\alpha$, $\beta$, $\gamma$ and $\delta$}.

\begin{figure}[!h]
	\includegraphics[width=8.5cm]{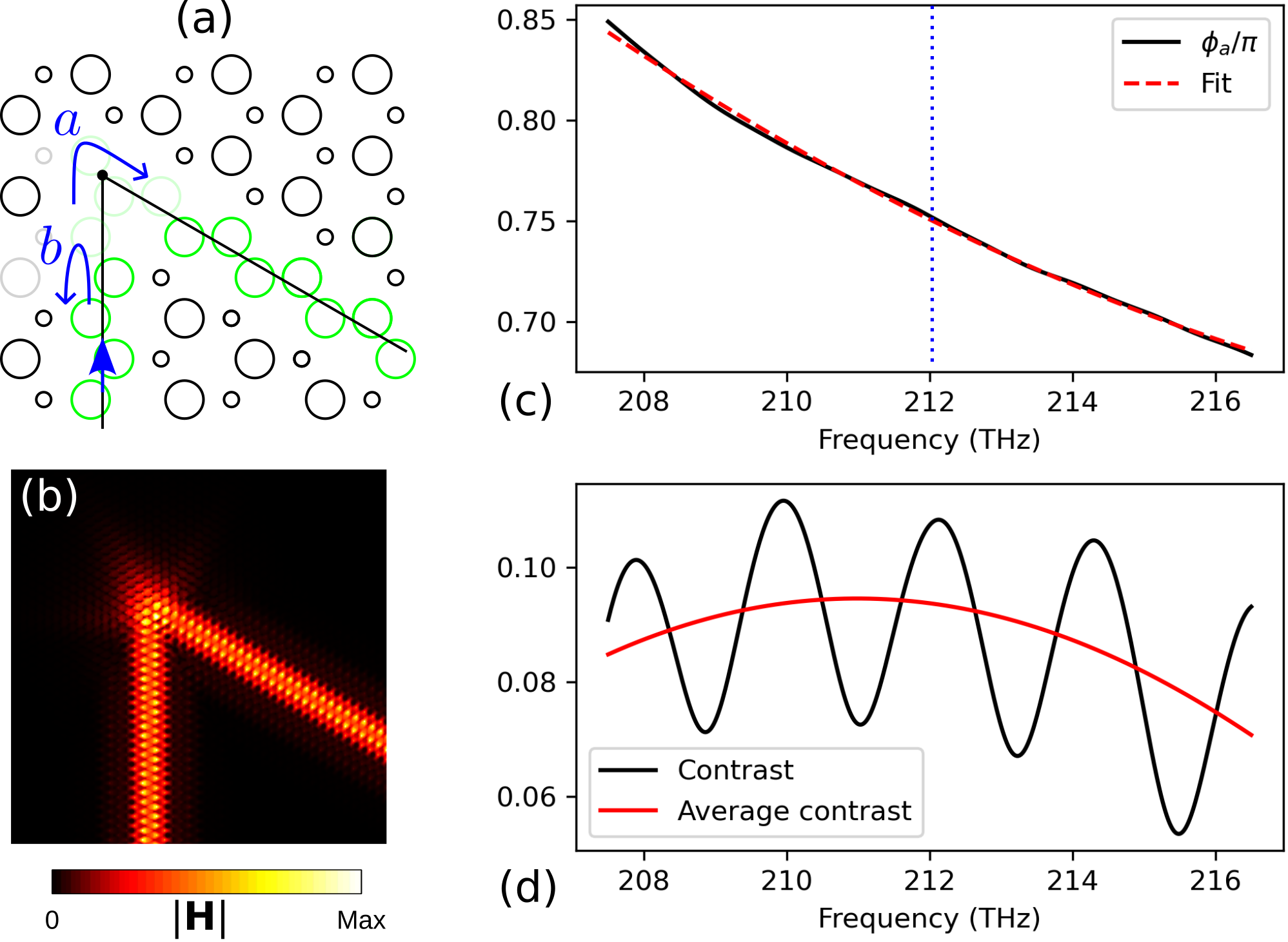}
	\caption{(a) Transmission, $a$, and reflection, $b$, coefficients on a corner. (b) Distribution of the magnetic field amplitude at $F=212$ THz. (c) Transmission phase $\phi_a$ normalized to $\pi$ computed from numerical simulations. (d) Contrast of the interference pattern along the vertical input edge of figure (a), black solid line, and average contrast, solid red line.}
	\label{fig:paramC}
\end{figure}

We now need to numerically evaluate \textcolor{black}{eight real} parameters: two for $\mathbf{M}_C$ and \textcolor{black}{six} for $\mathbf{M}_S$. To that purpose, reference points have to be defined in order to evaluate the phases of the matrices coefficients. Concerning the corner, the reference is taken as the intersection of the average lines of the bearded edges, as shown on Fig.~\ref{fig:paramC}(a). The transmission, $a=\exp{i\tau}\cos\sigma$, and reflection, $b=i\exp{i\tau}\sin\sigma$, coefficients have been numerically evaluated simulating the propagation of the edge mode along a single corner, whose magnetic field amplitude distribution is shown on Fig.~\ref{fig:paramC}(b). The phase $\phi_{a}=\tau$ of the transmission coefficient $a$ is obtained by comparing the phase of the transmitted mode to the phase of a mode propagating along a straight waveguide (see section 4 of the supplementary informations file). The numerical phase and a second order polynomial fit have been plotted on Fig.~\ref{fig:paramC}(c). The amplitude $\abs{b}$ of the reflection coefficient is directly given by the contrast of the interference pattern in the injection guide, which is observable, despite being weak, on the magnetic field distribution of Fig.~\ref{fig:paramC}(b). Figure \ref{fig:paramC}(d) shows the contrast as a function of the frequency, which presents a noticeable oscillation due to the interference with the wave which is slightly back-scattered on the PML. However, we can obtain a correct approximation of $\abs{b}$ by evaluating the average contrast, on the order of 8-9\% in amplitude. As $\phi_b=\phi_a+\pi/2$ and $\abs{a}=\sqrt{1-\abs{b}^2}$, the matrix $\mathbf{M}_C$ is fully determined, within the simulation uncertainties.

\begin{figure}[!h]
	\includegraphics[width=7cm]{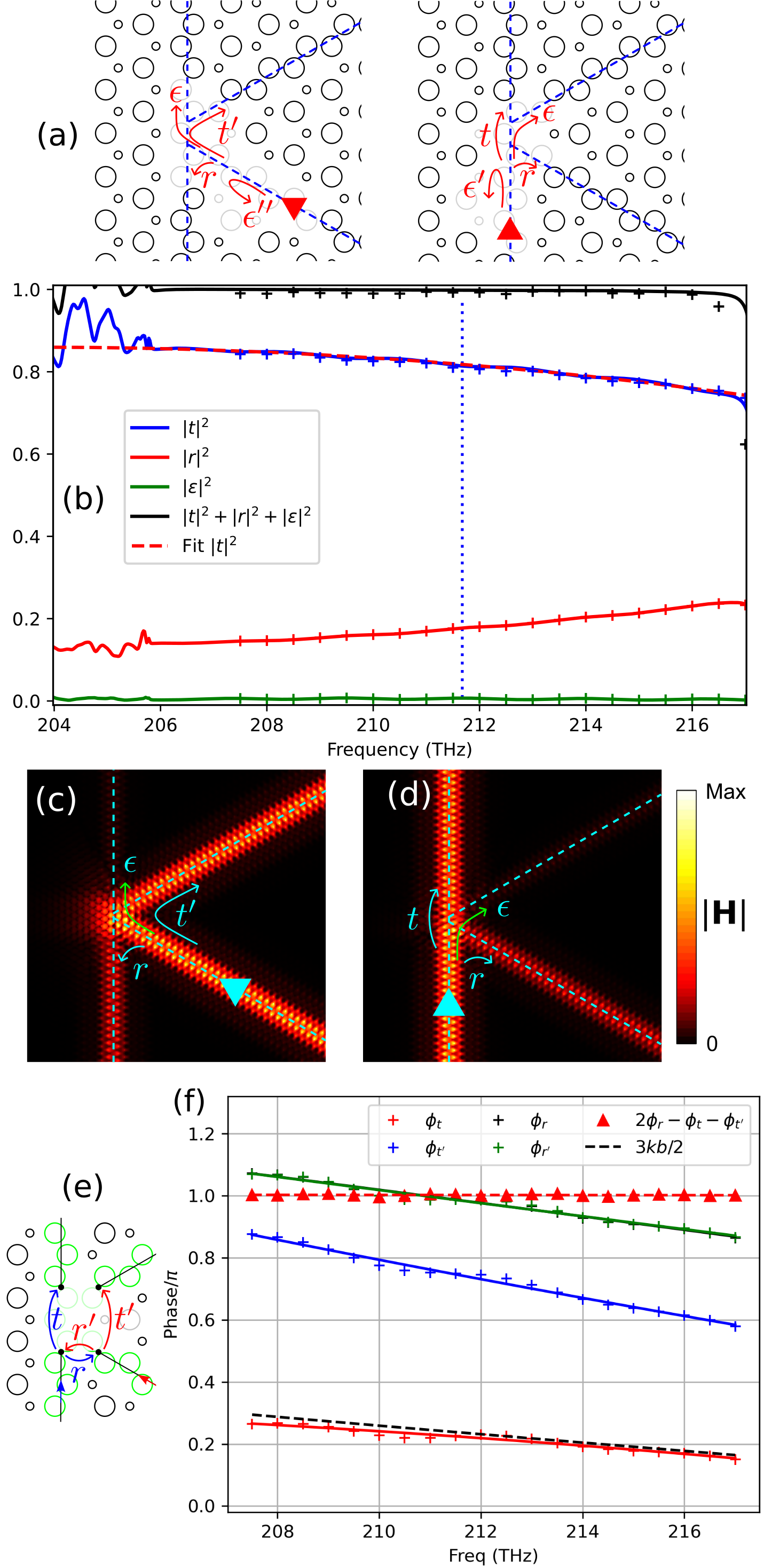}
	\caption{(a) Geometry of the splitter and coupling coefficients when addressed from an oblique (left) or a vertical (right) branch. (b) Evolution of the intensity of the waves propagating along each branch of the splitter. Solid lines (resp. crosses) correspond to the illumination along the vertical (resp. oblique) branch. (c) and (d) Distribution of the magnetic field amplitude respectively for oblique and vertical incidence. (e) Transmission and reflection coefficients across the splitter for vertical ($t$, $r$) and oblique ($t'$, $r'$) incidence. (f) Phases of the different coefficients across the splitter.}
	\label{fig:defcoeff}
\end{figure}

For the splitter, we have evaluated both the amplitudes and phases of the reflection ($r$) and transmission ($t$ and $t'$) coefficients defined in the semi-analytical model, see Eq.~\ref{eq:mk}. As indicated on Fig.~\ref{fig:defcoeff}(a), the four-branches splitter can be addressed either from one of the oblique (left panel) or the vertical (right panel) branches.
Following the definition of the coupling coefficients and taking the example of the oblique excitation, the wave can be transmitted with a coefficient $t'$ to the second oblique branch (called "transmission" branch), reflected with a coefficient $r$ into a topological mode with same helicity propagating along the downward vertical branch ("reflection" branch), coupled to a mode with opposite helicity with a coefficient $\epsilon$ propagating along the upward vertical branch ("forbidden" branch), or back-scattered into the mode with opposite helicity into the excitation branch with coefficient $\epsilon''$. Similarly in the second case, and with equivalent denomination, the wave coming from a vertical branch can be coupled to the transmission branch ($t$), the reflection branch (downward oblique branch, $r$), the forbidden branch (upward branch, $\epsilon$), or back-scattered ($\epsilon'$). As shown above, the coefficients $r$ and $\epsilon$ must be the same in both cases (excitation from a vertical or an oblique branch), but the transmission and back-scattering coefficients can be different. The evolutions with frequency of the squared modulus of the transmission, reflection and forbidden transmission coefficients are plotted on Fig.~\ref{fig:defcoeff}(b), in solid lines for the vertical excitation, and with colored crosses for the oblique excitation. First, we can verify that the transmission and the reflection coefficients have nearly equal values in both configurations. Second, it appears clearly that the coefficient $\epsilon$ is weak (see green solid line). For this reason, it cannot be evaluated by this method because the average value, below 1\% in intensity, could be related to the reflection of the transmitted and reflected waves on the PMLs surrounding the simulation domain. As a consequence, $\abs{\epsilon}^2$ can have any value between 0 and about 1\%. As numerically $\abs{t}^2+\abs{r}^2\approx 1$, $\epsilon'$ and $\epsilon''$ are confirmed to be, as $\epsilon$, much smaller than 1. In that case, as the coefficient $r$ is supposed to be the same in oblique or vertical excitation, energy conservation implies that $\abs{t}\approx\abs{t'}$, which is correctly reproduced by the numerical simulation. This tends to show that topological protection is mostly conserved at the splitter. The dashed line is a second order fit of $\abs{r}^2$, which varies between 84 and 76\%. The distribution of the magnetic field amplitude is plotted on Fig.~\ref{fig:defcoeff}(c) and (d) respectively for oblique and vertical excitation, at a frequency of 212 THz. We can visually verify the equality of the transmission and reflection coefficients. A very faint field can be distinguished along the forbidden channel, which may be again attributed to weak reflection of the transmitted and reflected fields on PMLs. As a conclusion, it is reasonable to consider \textit{as a first approximation} that the splitter conserves topological protection: an incident topological mode can be either coupled to the transmission channel (characterized by coefficients $t$ of $t'$), the reflection channel (coefficient $r$), but not to the forbidden channel, neither being back-reflected. Those results agree well with those obtained by Ma et al\cite{ma_topological_2019} for a similar splitter but with triangular holes. \textcolor{black}{In equation \ref{eq:mk}, it appears that perfect topological protection corresponds to $\phi=\phi'$ and $\gamma=\delta=0$, which we suppose in the next paragraph.}

In order to evaluate the phases of $t$, $t'$ and $r$, reference points have been defined as indicated on Fig.~\ref{fig:defcoeff}(e): the phase will be for each coefficient the phase difference of the topological mode between two points linked by the corresponding arrow.
Figure \ref{fig:defcoeff}(f) shows the frequency evolution of the four phases $\phi_{t}$, $\phi_{t'}$, $\phi_{r}$ and $\phi_{r'}$, normalized to $\pi$, corresponding to coefficients $t$, $t'$, $r$ and $r'$. 
We can first verify numerically that, as predicted by energy conservation and time-inversion symmetry, $\phi_{r}=\phi_{r'}$, and finally $r=r'$. The second point is that the phase $\phi_t$ for the transmission coefficient along the straight edge of the connection is equal, within the numerical uncertainties, to the propagation phase of the wave along the distance $3b_0/2$ between the two reference points of $t$. No additional phase is introduced by the presence of the nearby oblique edges. Next, the three phases $\phi_{t}$, $\phi_{t'}$ and $\phi_{r}$ are related to $\alpha$ and $\beta$ ($\gamma=\rho=0$) through:
\begin{align*}
    &\alpha+\beta+\pi/2=\phi_t\\
    &\alpha-\beta+\pi/2=\phi_{t'}\\
    &\alpha=\phi_r
\end{align*}
which leads to $\pi\qty[2\pi]=2\phi_{r}-\phi_{t}-\phi_{t'}$. We can see on Fig.~\ref{fig:paramC}(f) that this relation is very well verified numerically. As we suppose for now that the splitter preserves topological protection, the three remaining coefficients $\epsilon$, $\epsilon'$ and $\epsilon''$ are taken as 0.

\section{Comparison with FEM simulations}
The semi-analytical modelization of the triangular resonators is realized using a coupled wave approach, whose details are given in section 5 of the supplementary informations file. The principle is summarized on Fig.~S6, which shows how $C$ and $S$ points are connected through segments of different lengths. Due to the choice of reference points for the splitter, the lengths of the different triangle edges are slightly different ($L_0=28.5\,b_0$, $L_1=27\,b_0$ and $L_2=13.5\,b_0$).

\begin{figure}[!h]
	\includegraphics[width=8cm]{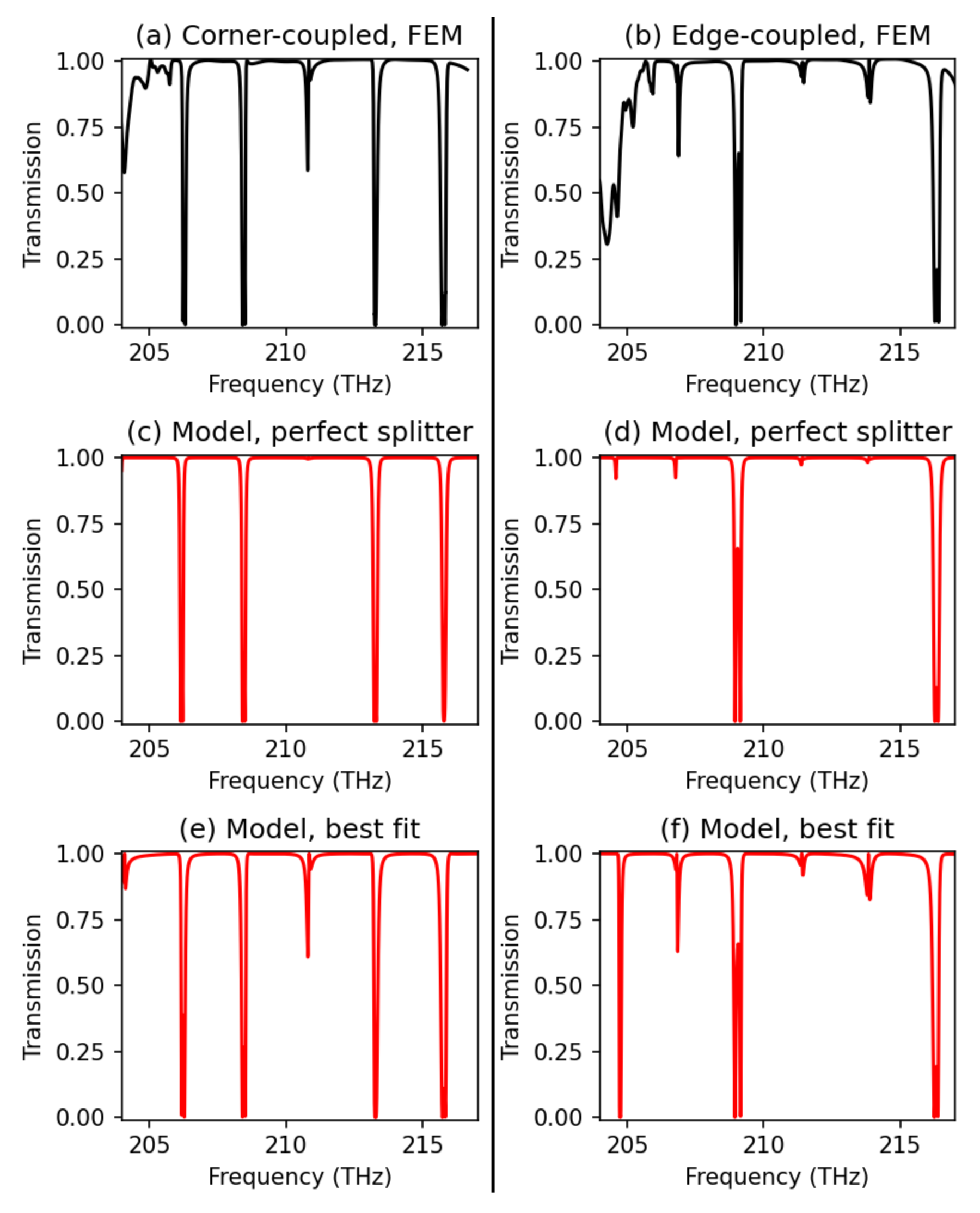}
	\caption{Comparison between the transmission spectra obtained from finite elements simulations \textcolor{black}{and the semi-analytical model for the, (a,c,e), corner-addressed and, (b,d,f), edge-addressed cavity. The splitter is either supposed to be perfectly topology-protected, (c,d), or to present small breakdown of topological protection, (e,f).}}
	\label{fig:transmissionref}
\end{figure}

The transmission spectra can be computed for triangular resonators either addressed from the corner or the edge. For the resonators of Fig.~\ref{fig:meta}, the spectra obtained by FEM simulations (Fig.~\ref{fig:transmissionref}\textcolor{black}{(a,b)}) compare very correctly with the semi-analytical model (bottom red lines) \textcolor{black}{even if the splitter is supposed to perfectly preserve topological protection, Fig.~\ref{fig:transmissionref}(c,d)}. Note that a slight adjustment of the edge mode wavevector (by 0.06\%) has been realized to have a slightly better agreement on the resonance frequencies. The frequencies and alternation of the split resonances ($T=0$) and anti-resonances ($T=1$) are well reproduced for both corner- or edge-addressed resonators, and the FWHM are comparable.
The main difference is the profiles close to the anti-resonances, which are barely visible \textcolor{black}{on figures (c) and (d)}, but much more pronounced in the numerical simulation, with a strong asymmetry. This fact \textcolor{black}{has then to} be related to a breakdown of the topological protection at the splitter. \textcolor{black}{Despite the fact that the }parameter space of the system is large, with eight free real parameters, \textcolor{black}{it is possible to adjust the remaining small coefficients ($\epsilon$, $\epsilon'$ and $\epsilon''$) by fitting each resonance profiles in a narrow frequency region around them. This method allows reaching a much better agreement, as shown on Fig.~\ref{fig:transmissionref}(e,f). Indeed, all the missing features are now recovered, concerning asymmetry of the split resonances and the exact profile of the anti-resonances. Figure S7 shows the estimated frequency-evolution of the squared amplitude of the eight coefficients corresponding to the corner and splitter scattering matrices. As supposed, the largest coefficient at the origin of the pseudo-spin conversion is $b$, the reflection on a corner, but the coefficients $\epsilon$, $\epsilon'$ and $\epsilon''$ are finally comparable. For example, at $F=212$ GHz, we have $\abs{a}^2=0.99$ and $\abs{b}^2=0.01$ for the corner, and $\abs{t}^2=0.784$, $\abs{t'}^2=0.783$, $\abs{r}^2=0.209$, $\abs{\epsilon}^2=5.\,10^{-3}$, $\abs{\epsilon'}^2=1.3\,10^{-3}$ and $\abs{\epsilon''}^2=2.2\,10^{-3}$ for the splitter.} Hence, the loss of topological protection is evaluated to be \textcolor{black}{more than two} times larger in power at the sharp corners of the triangular resonator (back-scattering) than through the splitter (back-scattering and forbidden transmission).

As a last study, we \textcolor{black}{propose to assess the robustness of the semi-analytical approach by comparisons with full numerical simulations of larger and more complex resonators. They consist in three fractal-inspired structures based on the Sierpi\'nski triangle construction, presented on Fig.~\ref{fig:transmissionrefdec}. The first structure, Fig.~\ref{fig:transmissionrefdec}(a), is simply a corner-addressed triangle whose edges have a length of 51 periods. The second, Fig.~\ref{fig:transmissionrefdec}(b), is the first iteration of the Sierpi\'nski construction, and is composed of four triangles with an edge-length of 26 periods. 
\begin{figure}[!h]
	\includegraphics[width=15cm]{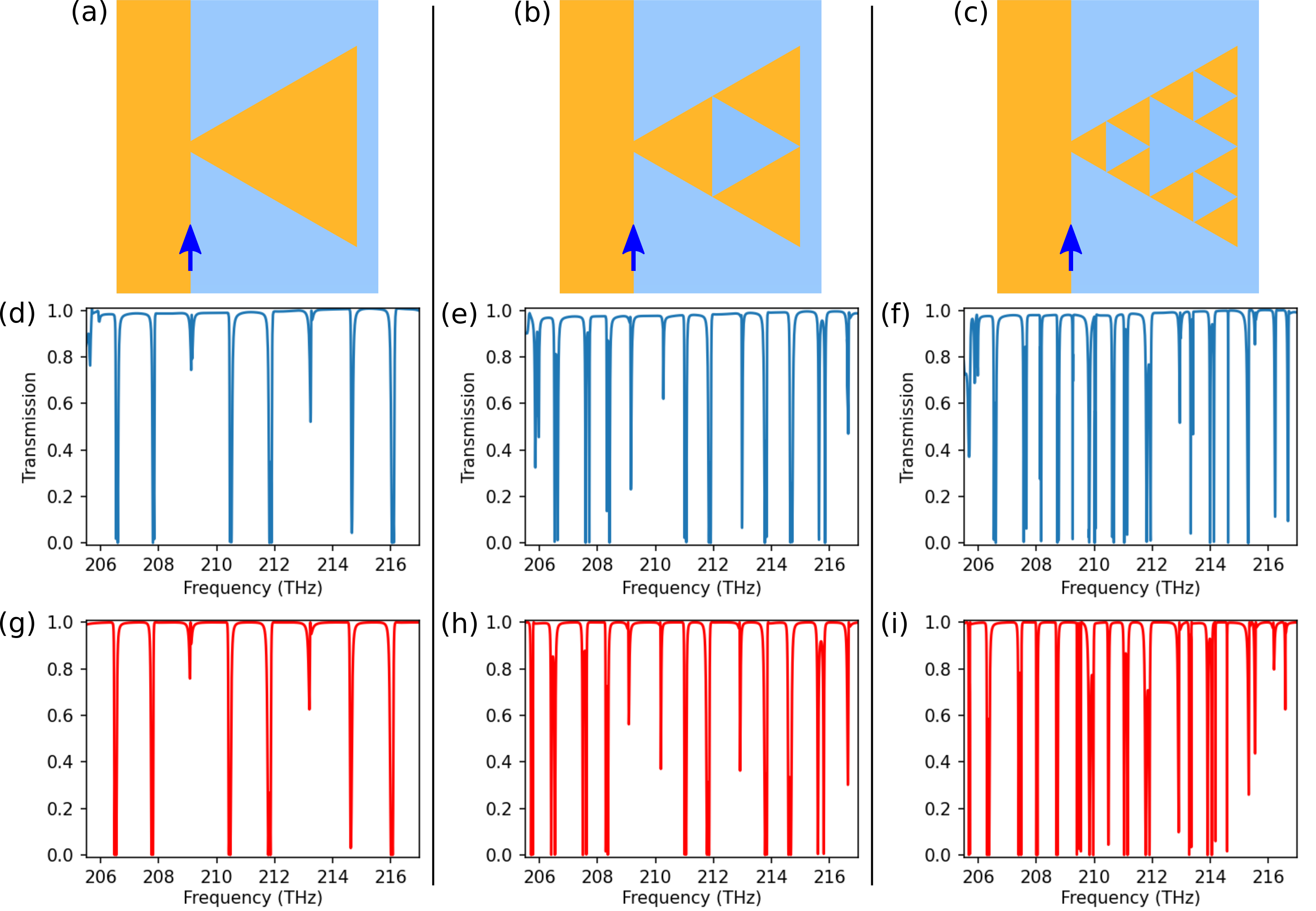}
	\caption{\textcolor{black}{Comparison between the full numerical simulation and the semi-analytical model for the three first iteration of the Sierpi\'nski triangle construction, (a,b,c). (d,e,f) Corresponding transmission spectra computed with the finite-element-method. (g,h,i) Transmission spectra calculated with the semi-analytical model.}}
	\label{fig:transmissionrefdec}
\end{figure}
The last resonator, Fig.~\ref{fig:transmissionrefdec}(c), is the second iteration, and is an assembly of one triangle with 26-periods-long edges and 12 triangles with 13-periods-long edges. The transmission spectrum of the simple triangle is very similar to the resonator previously discussed, and show a similar regular alternation of split-resonances and anti-resonances, however separated by a smaller frequency interval as the cavity is larger. The comparison between the finite-element simulation, Fig.~\ref{fig:transmissionrefdec}(d), and the semi-analytical model, Fig.~\ref{fig:transmissionrefdec}(g), is again very good. For the second structure, more interference paths are possible in the resonator: the simulated transmission spectrum is richer and does not present the same regular pattern obtained for the simple triangle, Fig.~\ref{fig:transmissionrefdec}(e). However, the agreement with the semi-analytical model is still good, Fig.~\ref{fig:transmissionrefdec}(h): single and split-resonances are recovered at the same frequencies, but small differences in amplitude are observed for shallower resonances, close to 209, 210, 213 and 217 THz. The semi-analytical model starts to differ more significantly from the numerical simulations for the second iteration. The numerical transmission, Fig.~\ref{fig:transmissionrefdec}(f), presents consistently more resonances, single or split in two or more peaks. It appears that the number, positions and amplitudes of those resonances do not coincide as correctly with the semi-analytical model, even if some similarities are observed, for example in term of the density of resonances as a function of the frequency, Fig.~\ref{fig:transmissionrefdec}(i). Two main explanations can be proposed. First, as the interference paths are more complex, small errors in the estimations of the matrices coefficients have more significant impact on the transmission. Second, as the edges are shorter, we can expect that the model becomes less valid, as the edge-mode, which has a lateral extend, can directly tunnel laterally to adjacent edges across the lattice. An example of field distribution is shown on Fig.~S8, for the resonance at F=214.61 GHz. Despite those limits, our results are encouraging and show that such a semi-analytical approach can be employed to predict, with low computer resources, the propagation of the electromagnetic signal along a complex and extended topological circuit.}

\section*{Conclusion}
In this work, we have investigated, using a full numerical approach based on the finite elements method and a semi-analytical model relying on an exact parameterization of scattering matrices at splitters and corners, the breakdown of topological protection evidenced by the resonance properties of valley topological triangular resonators. In resonators, perfect topological protection implies a flat transmission band, then any resonance feature must result from a back-scattering or forbidden transmission between waves with opposite helicity occurring at particular points of the system (corners, splitters...). In our system, we have demonstrated that the split-resonances, together with anti-resonances, must be mainly attributed to back-scattering at corners of the triangular cavity with a lower but comparable contribution of the splitter. \textcolor{black}{Quantitatively, and for the considered valley topological crystal, the amount of power back-scattered at corners is about 1\% of the incident edge-mode, while the backscattering and forbidden transmission at the splitter is lower than 0.5\%. We have then demonstrated, by simulations of fractal-inspired larger resonators, that our semi-analytical approach can be employed for fast and reliable predictive simulations of larger and more complex topological systems, if however the length of the edges composing the circuit are not too short (larger than 13 periods in our study) in order to avoid unwanted tunneling between close edges through the photonic lattice.} We believe that the proposed methodology can be applied to different geometry of topological photonic devices in order to evaluate the quality of the topological protection depending on the shape (triangular vs circular) of air holes, edges (bearded or zigzag holes), or configuration of splitters (four or six branches) and corners, which is a crucial point in order to design photonic devices which gather compactness and low losses made possible through topological conduction of electromagnetic waves.

\paragraph{Acknowledgements}
This work was supported by the Horizon-RIA action project "Magnific" (101091968).

\providecommand{\latin}[1]{#1}
\makeatletter
\providecommand{\doi}
  {\begingroup\let\do\@makeother\dospecials
  \catcode`\{=1 \catcode`\}=2 \doi@aux}
\providecommand{\doi@aux}[1]{\endgroup\texttt{#1}}
\makeatother
\providecommand*\mcitethebibliography{\thebibliography}
\csname @ifundefined\endcsname{endmcitethebibliography}
  {\let\endmcitethebibliography\endthebibliography}{}



\section*{Supplementary material}

# Supplementary Informations for Quantifying topological protection in valley photonic crystals using triangular resonators

Gaëtan Lévêque,*,† Yan Pennec,† Pascal Szriftgiser,‡ Alberto Amo,‡ and

Alejandro Martínez¶

†Institut d'Electronique, de Micro-électronique et de Nanotechnologie (IEMN, CNRS-8520), Cité Scientifique, Avenue Poincaré, 59652 Villeneuve d'Ascq, France

‡Université de Lille, CNRS, UMR 8523-PhLAM-Physique des Lasers Atomes et Molécules, F-59000 Lille, France

¶1Nanophotonics Technology Center, Universitat Politècnica de València, Camino de Vera s/n, 46022 Valencia, Spain

E-mail: gaetan.leveque@univ-lille.fr

# 1 Berry curvature simulations

In order to evaluate the Berry phase, the first Brillouin zone (FBZ) is discretized into a 70 × 70 grid matching the FBZ geometry (not shown here). With Matlab, thanks to the plane wave expansion method, the periodic part of the $H_z$ field component is evaluated over each vertice of the grid. The berry curvature is reconstructed by computing the accumulated phase around the closed loop made of consecutive points of the discretized grid, see reference [1] for more details. Since the first band is isolated from all other bands, its Berry curvature

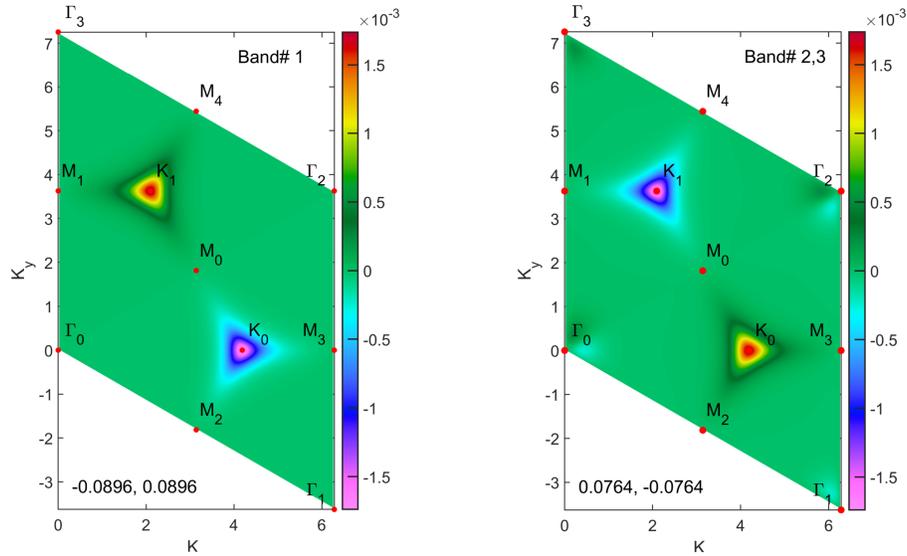

Figure S1: Numerical computation of the Berry curvature.

can be computed on its own (Figure S1, left). Next, by integrating the Berry curvature inside the triangle delimited by points $\Gamma_0$, $\Gamma_2$ and $\Gamma_3$ on the one hand, and points $\Gamma_0$, $\Gamma_1$ and $\Gamma_2$ on the other hand, one can recover the valley Chern numbers associated with the first band, i.e. -0.090 and + 0.090. However, bands 2 and 3 are degenerated, i.e. they share identical frequencies at the same location (Figure S1, right). They are however isolated from all other bands. The Berry curvature must then be computed as a whole for those two bands. This is performed following the non-Abelian Berry connection generalization,[1,2] resulting into +0.076 and -0.076 valley Chern numbers.

# 2 Comsol simulations

Numerical simulations were realized using the finite elements software Comsol Multiphysics. A typical setup is presented on figure S2. The simulation domain, in light gray, comprises the resonator together with the input and output waveguides, emphasized with red lines. Note that the topological photonic crystal extends within the PMLs (Perfectly Matched Layers) surrounding the physical domain: this configuration has been tested and chosen because it lead to weak back-reflection. PMLs have been set to cartesian, except for the two tilted ones

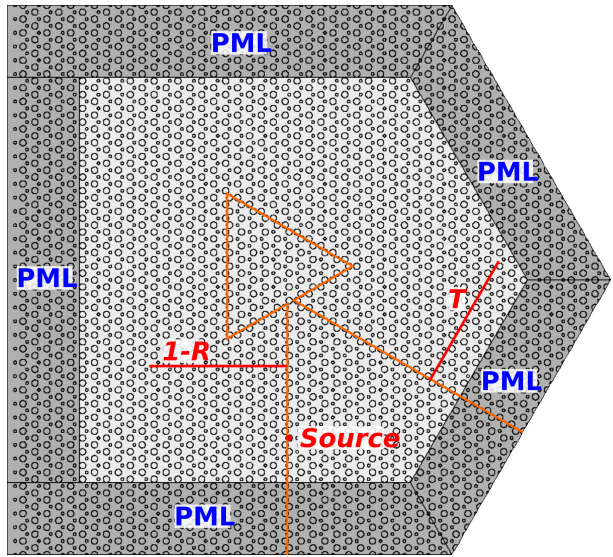

Figure S2: Comsol configuration.

where the direction of attenuation has to be chosen by hand. The input intensity $1 - R$ has been evaluated by integration of the Poynting vector along the horizontal line behind the source, only on the left side to avoid perturbation due to the output waveguide. Similarly, the transmission $T$ is evaluated along a line placed just before the PMLs corresponding to the output waveguide. The source is a rotating dipole localized at the center of one of the large holes of the topological waveguide.

# 3 Semi-analytical model: coupling matrices

We derive in this section the analytical expressions of the scattering matrices between in and out-coming topological edge modes at either the corner of the linear waveguide (two inputs and two outputs) or a splitter (four inputs and four outputs). The calculation is based on energy conservation, time-reversal invariance and mirror symmetries of the configuration.

## 3.1 Scattering matrix of a corner

A simple corner, in the most general situation, behaves like a coupler between one of the two incident edge modes $A^+$ and $B^+$ with the corresponding transmitted and reflected modes, $A^-$ and $B^-$, see Fig. S3. The process is linear and obeys the following relation:

$$\begin{bmatrix} A^- \\ B^- \end{bmatrix} = \mathbf{M}_C \begin{bmatrix} A^+ \\ B^+ \end{bmatrix} = \begin{bmatrix} t_1 & r_2 \\ r_1 & t_2 \end{bmatrix} \begin{bmatrix} A^+ \\ B^+ \end{bmatrix}$$

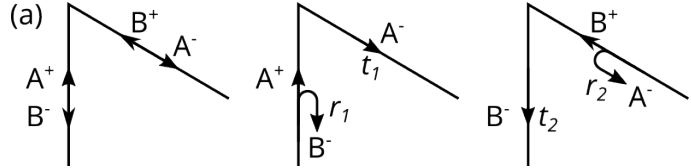

Figure S3: Definition of the wave amplitudes and coupling coefficients for a corner.

Energy conservation implies that:

$$\left|A^-\right|^2 + \left|B^-\right|^2 = \left|A^+\right|^2 + \left|B^+\right|^2, \ \forall A^+, B^+$$

The matrix $\mathbf{M}_C$ is then unitary:

$$\mathbf{M}_C^\dagger \mathbf{M}_C = \mathbf{1}.$$

Additionally, time-reversal symmetry implies that :

$$\begin{bmatrix} B^+ \\ A^+ \end{bmatrix}^* = \mathbf{M}_C \begin{bmatrix} B^- \\ A^- \end{bmatrix}^* \Leftrightarrow \mathbf{S}_C \begin{bmatrix} A^+ \\ B^+ \end{bmatrix}^* = \mathbf{M}_C \mathbf{S}_C \begin{bmatrix} A^- \\ B^- \end{bmatrix}^* \quad \text{with} \quad \mathbf{S}_C = \begin{bmatrix} 0 & 1 \\ 1 & 0 \end{bmatrix}$$

which, combined with unitarity, implies $\mathbf{M}_C = \mathbf{S}_C \mathbf{M}_C^T \mathbf{S}_C$, and $t_1 = t_2$:

$$\mathbf{M}_C = \begin{bmatrix} t & r_1 \\ r_2 & t \end{bmatrix}$$

The unitarity condition imposes:
$$|r_1|^2 + |t|^2 = 1$$
$$|r_2|^2 + |t|^2 = 1 \qquad (1)$$
$$r_1 t^* + t r_2^* = 0$$

The two first equations impose $|r_1| = |r_2| = R$. Let us write $t = T \exp(i\tau)$, $r_1 = R \exp(i\rho_1)$ and $r_2 = R \exp(i\rho_2)$. We have $R^2 + T^2 = 1$, and:

$$RT \left( e^{i(\rho_1 - \tau)} + e^{i(\tau - \rho_2)} \right) = 0 \implies \rho_1 - \tau = \tau - \rho_2 + \pi + m 2\pi, \ m \in \mathbb{Z}$$

or:

$$\rho_1 - \tau - \pi/2 + \rho_2 - \tau - \pi/2 + m 2\pi = 0$$

Writing:

$$r_1 = R \exp(i(\rho_1 - \tau - \pi/2 + \tau + \pi/2))$$
$$r_2 = R \exp(i(\rho_2 - \tau - \pi/2 + \tau + \pi/2))$$
$$= R \exp(i(-(\rho_1 - \tau - \pi/2) + \tau + \pi/2)),$$

5

we pose $\rho = \rho_1 - \tau - \pi/2$ and we obtain $r_1 = iR\exp(i(\rho+\tau))$ and $r_2 = iR\exp(i(-\rho+\tau))$. Finally, the $\mathbf{M}_C$ matrix reads:

$$\mathbf{M}_C = e^{i\tau}\begin{bmatrix} \cos\sigma & i\sin\sigma e^{i\rho} \\ i\sin\sigma e^{-i\rho} & \cos\sigma \end{bmatrix} \quad (2)$$

where we have introduced the angle $\sigma$ such that $T = \cos\sigma$ and $R = \sin\sigma$. We end up with three degrees of freedom. Note that the eigenvalues of $\mathbf{M}_C$ are $\lambda^\pm = \exp\{i(\tau\pm\sigma)\}$. Assuming symmetry of the junction, we can as well impose $r_1 = r_2 \Rightarrow \rho = 0$, and, with $\sigma \in [0,\pi]$ such as $R = \sin\sigma$ and $T = \cos\sigma$:

$$\mathbf{M}_C = e^{i\tau}\begin{bmatrix} \cos\sigma & i\sin\sigma \\ i\sin\sigma & \cos\sigma \end{bmatrix}$$

For perfect topological protection, $\sigma = 0$ and $\mathbf{M}_C$ is diagonal.

## 3.2 Scattering matrix of a splitter

Let us consider now the splitter. Following Fig. S4, four outputs, $\mathbf{X}^+ = [A^+, B^+, C^+, D^+]^T$, are now connected to four inputs, $\mathbf{X}^- = [A^-, B^-, C^-, D^-]^T$. We must then extend the $2\times 2$ to a $4\times 4$ matrix with the following shape:

$$\mathbf{X}^- = \mathbf{M}_S\mathbf{X}^+ \quad \Leftrightarrow \quad \begin{bmatrix} A^- \\ B^- \\ C^- \\ D^- \end{bmatrix} = \begin{bmatrix} r_1 & \epsilon_2 & \epsilon'_3 & t_4 \\ \epsilon_1 & r_2 & t_3 & \epsilon'_4 \\ \epsilon'_1 & t_2 & r_3 & \epsilon_4 \\ t_1 & \epsilon'_2 & \epsilon_3 & r_4 \end{bmatrix}\begin{bmatrix} A^+ \\ B^+ \\ C^+ \\ D^+ \end{bmatrix}$$

The mirror-symmetry of the corner, with respect to the dashed line on Fig. 4, leads to

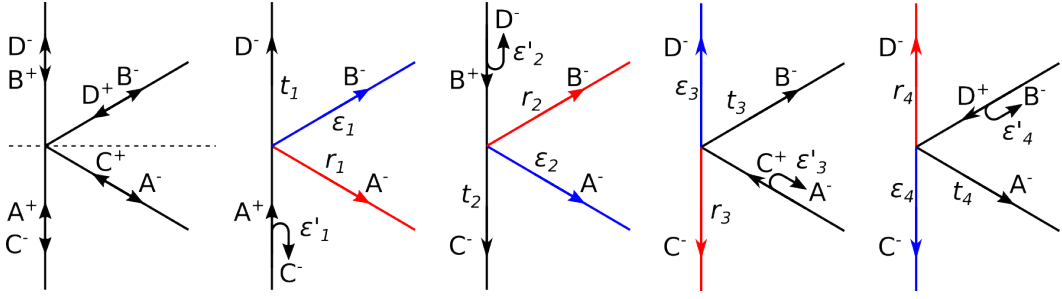

Figure S4: Definition of the wave amplitudes and coupling coefficients for a splitter.

$\alpha_1 = \alpha_2$ and $\alpha_3 = \alpha_4$ where $\alpha = r, t, \epsilon, \epsilon'$. Additionally, time-reversal symmetry imposes:

$$\begin{bmatrix} C^+ \\ D^+ \\ A^+ \\ B^+ \end{bmatrix}^* = \mathbf{M}_S \begin{bmatrix} C^- \\ D^- \\ A^- \\ B^- \end{bmatrix}^* \Leftrightarrow \mathbf{S}_S(\mathbf{X}^+)^* = \mathbf{M}_S \mathbf{S}_S(\mathbf{X}^-)^*, \quad \mathbf{S}_S = \begin{bmatrix} 0 & 0 & 1 & 0 \\ 0 & 0 & 0 & 1 \\ 1 & 0 & 0 & 0 \\ 0 & 1 & 0 & 0 \end{bmatrix}$$

Combined with unitarity of $\mathbf{M}_S$, that last relation leads to:

$$\mathbf{M}_S = \mathbf{S}_S \mathbf{M}_S^T \mathbf{S}_S.$$

By comparison of both members of that equation, we finally see that $r_1 = r_2 = r_3 = r_4$ and $\epsilon_1 = \epsilon_2 = \epsilon_3 = \epsilon_4$. The matrix $\mathbf{M}_S$ can then be expressed under the following form:

$$\mathbf{M}_S = \begin{bmatrix} r & \epsilon & \epsilon'' & t' \\ \epsilon & r & t' & \epsilon'' \\ \epsilon' & t & r & \epsilon \\ t & \epsilon' & \epsilon & r \end{bmatrix}$$

As $\mathbf{M}_S$ is unitary, it has four eigenvalues with unit modulus. We can exploit this property in order to find the full parameterization of the matrix. For that, we notice that the eigenvectors must be either symmetric or anti-symmetric with respect to the line of symmetry of the splitter (dotted line on Fig. 4).

A set of symmetric vectors is $\mathbf{A}_1 = [1, 1, 0, 0]^T$ and $\mathbf{A}_2 = [0, 0, 1, 1]^T$, in which basis the matrix $\mathbf{M}_S$ reduces to:

$$\mathbf{M}_S^A = \begin{bmatrix} r + \epsilon & t' + \epsilon'' \\ t + \epsilon' & r + \epsilon \end{bmatrix}$$

As this matrix is unitary, it has the form given by equation 2, and :

$$r + \epsilon = \cos \phi e^{i\alpha}, \qquad t' + \epsilon'' = i \sin \phi e^{i(\alpha+\beta)}, \qquad t + \epsilon' = i \sin \phi e^{i(\alpha-\beta)}$$

Similarly, a set of anti-symmetric vectors is $\mathbf{B}_1 = [1, -1, 0, 0]^T$ and $\mathbf{B}_2 = [0, 0, -1, 1]^T$, in which basis the matrix $\mathbf{M}_S$ reduces to:

$$\mathbf{M}_S^B = \begin{bmatrix} r - \epsilon & t' - \epsilon'' \\ t - \epsilon' & r - \epsilon \end{bmatrix}$$

which leads to:

$$r - \epsilon = \cos \phi' e^{i\alpha'}, \qquad t' - \epsilon'' = i \sin \phi' e^{i(\alpha'+\beta')}, \qquad t - \epsilon' = i \sin \phi' e^{i(\alpha'-\beta')}$$

Finally, we obtain:

$$r = \frac{1}{2}\left(c_\phi e^{i\alpha} + c_{\phi'} e^{i\alpha'}\right) \qquad\qquad \epsilon = \frac{1}{2}\left(c_\phi e^{i\alpha} - c_{\phi'} e^{i\alpha'}\right)$$
$$t' = \frac{i}{2}\left(s_\phi e^{i(\alpha+\beta)} + s_{\phi'} e^{i(\alpha'+\beta')}\right) \qquad\qquad \epsilon'' = \frac{i}{2}\left(s_\phi e^{i(\alpha+\beta)} - s_{\phi'} e^{i(\alpha'+\beta')}\right)$$
$$t = \frac{i}{2}\left(s_\phi e^{i(\alpha-\beta)} + s_{\phi'} e^{i(\alpha'-\beta')}\right) \qquad\qquad \epsilon' = \frac{i}{2}\left(s_\phi e^{i(\alpha-\beta)} - s_{\phi'} e^{i(\alpha'-\beta')}\right)$$

with $c_\phi = \cos \phi$, $c_{\phi'} = \cos \phi'$, $s_\phi = \sin \phi$ and $s_{\phi'} = \sin \phi'$.

After some change of notation, we obtain:

$$\mathbf{M}_S = \frac{e^{i\alpha}}{2} \times$$

$$\begin{bmatrix} (c_\phi\, e^{i\rho} + c_{\phi'}\, e^{-i\rho}) & (c_\phi\, e^{i\rho} - c_{\phi'}\, e^{-i\rho}) & i(s_\phi\, e^{-i\delta} - s_{\phi'}\, e^{i\delta})e^{-i\beta} & i(s_\phi\, e^{-i\delta} + s_{\phi'}\, e^{i\delta})e^{-i\beta} \\ (c_\phi\, e^{i\rho} - c_{\phi'}\, e^{-i\rho}) & (c_\phi\, e^{i\rho} + c_{\phi'}\, e^{-i\rho}) & i(s_\phi\, e^{-i\delta} + s_{\phi'}\, e^{i\delta})e^{-i\beta} & i(s_\phi\, e^{-i\delta} - s_{\phi'}\, e^{i\delta})e^{-i\beta} \\ i(s_\phi\, e^{-i\gamma} - s_{\phi'}\, e^{i\gamma})e^{i\beta} & i(s_\phi\, e^{-i\gamma} + s_{\phi'}\, e^{i\gamma})e^{i\beta} & (c_\phi\, e^{i\rho} + c_{\phi'}\, e^{-i\rho}) & (c_\phi\, e^{i\rho} - c_{\phi'}\, e^{-i\rho}) \\ i(s_\phi\, e^{-i\gamma} + s_{\phi'}\, e^{i\gamma})e^{i\beta} & i(s_\phi\, e^{-i\gamma} - s_{\phi'}\, e^{i\gamma})e^{i\beta} & (c_\phi\, e^{i\rho} - c_{\phi'}\, e^{-i\rho}) & (c_\phi\, e^{i\rho} + c_{\phi'}\, e^{-i\rho}) \end{bmatrix}$$

(3)

with $\rho = -(\gamma + \delta)/2$. There are then 6 degrees of freedom.

If the topological protection is perfect ($\epsilon = \epsilon' = \epsilon'' = 0$), we have $\phi' = \phi$ and $\gamma = \delta = \rho = 0$, which gives:

$$\mathbf{M}_S^0 = e^{i\alpha} \begin{bmatrix} \cos\phi & 0 & 0 & i\sin\phi\, e^{-i\beta} \\ 0 & \cos\phi & i\sin\phi\, e^{-i\beta} & 0 \\ 0 & i\sin\phi\, e^{i\beta} & \cos\phi & 0 \\ i\sin\phi\, e^{i\beta} & 0 & 0 & \cos\phi \end{bmatrix}$$

Finally, starting from the expressions of $\mathbf{M}_S^A$ and $\mathbf{M}_S^B$, we show easily that the four eigenvalues of $\mathbf{M}_S$ are:

$$\lambda_A^\pm = e^{i(\alpha+\rho\pm\phi)}$$

$$\lambda_B^\pm = e^{i(\alpha+\rho\pm\phi')}$$

# 4 Estimation of the corner $a$ and $b$ coefficients

The transmission, $b$, and reflection, $a$, coefficients of a single corner have been estimated by comparing the component $H_z$ of the magnetic field computed either along a straight (solid line on Fig. S5(a), left) or curved (solid line on Fig.S5(a), right) topological edge, both lines starting just beside the source (red dot). The real part and the amplitude of $H_z$ are plotted respectively on Fig. S5(b) and (c), for $F = 212.3$ THz.

The phase of the field after the corner in the curved waveguide differs by the transmission phase $\phi_a$ from the field propagating along the straight edge, as can be seen on Fig. S5(b). The phase $\phi_a$ can then simply be extracted from the complex value of $H_z$ along the curved edge by comparison with the straight edge.

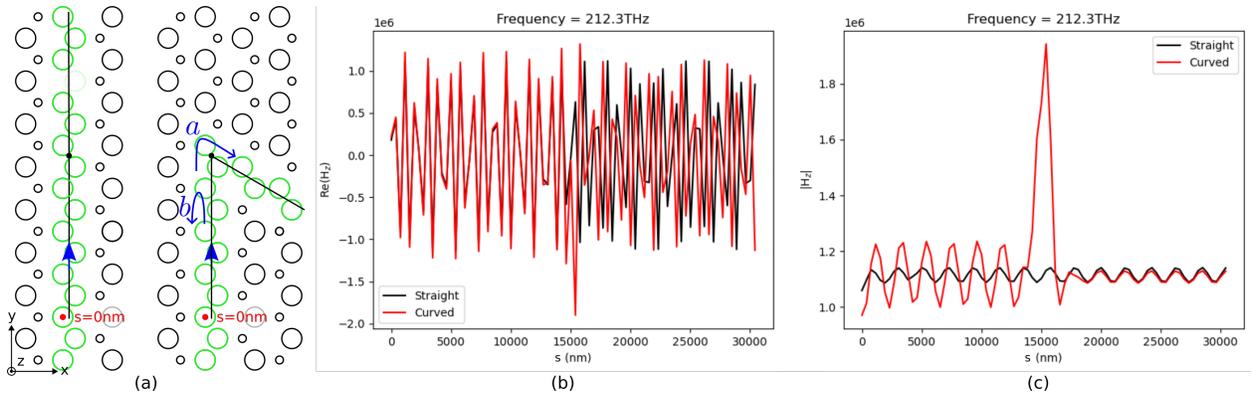

Figure S5: (a) Straight (left) and curved (right) edges along which the $z$ component of the magnetic field has been computed. (b) Real part of $H_z$ along the straight or curved path, as a function of the curvilinear position $s$ measured from the position of the source (red dot). (c) Same as (b) for the amplitude of $H_z$.

Concerning the amplitude of $b$, we can see on Fig. S5(c) that the reflection on the corner induces a noticeable amplitude modulation of the field before the corner ($s < 15\mu$m). A smaller modulation is seen on the field propagating along the straight edge and the field transmitted though the corner, which is related to a small reflection on the PMLs. The

superposition $H_c$ of the incident and reflected fields with amplitude $H_0$ reads:

$$H_c(y) = H_0[\exp\{ik(y-y_0)\} + b\exp\{-ik(y-y_0)\}]$$
$$\Rightarrow |H_c|(y) = |H_0|\sqrt{1 + |b|^2 + 2|b|\cos\{2k(y-y_0) - \phi_b\}}$$

where $y_0$ is the position of the corner and $k$ the wavevector of the edge mode. We can verify that the contrast in amplitude is $|b|$. Numerically, the simulation data corresponding to Fig. S5(c) have been fitted using a function $f(y,r,s,t,u) = (r + s\sin\{ty + u\})^{1/2}$, leading to:

$$|b| = \frac{\sqrt{r+s} - \sqrt{r-s}}{\sqrt{r+s} + \sqrt{r-s}}$$

As explained in the article, the extracted coefficient contains the contribution from PMLs, which is the reason why the average value of $|b|$ on the frequency window has been retained. As $\phi_a$ and $|b|$ are known, the reflection and transmission coefficients are respectively given by:

$$a = \sqrt{1-|b|^2}\exp\{i\phi_a\}$$
$$b = i|b|\exp\{i\phi_a\}$$

11

# 5 Semi-analytical model: cavities description

We establish in this section the matrix formalism used to model analytically the transmission across two types of triangular cavities. Both are excited by a waveguide linked to the cavity by a splitter, either directly at the tip (Fig. 6(a)), or at the edge (Fig. 6(b)). Notice that the different segments between corners $C$ and splitter $S$ can have different lengths, even in the case of the tip-coupled cavity, due to the exact position of air holes forming the splitter. For those reasons, $C - C$ and $S - C$ lengths are respectively written $L_0$ and $L_1$, while in the edge-coupled cavity we need to introduce a third length $L_2$ (close to $L_0/2$) for the $S - C$ segments.

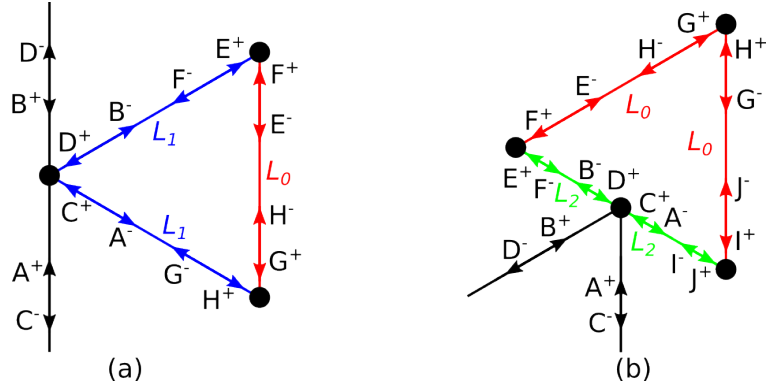

Figure S6: Simplified description of tip- and edge-coupled cavities in the semi-analytical model. The waves crossing the splitters and corners are connected by scattering matrices at nodes and propagation phases along the segments.

In both cases, the incoupling field have the form:

$$\mathbf{X}^+ = \begin{bmatrix} A^+ \\ B^+ \\ \mathbf{Y}^+ \end{bmatrix} \qquad A^+ = 1,\ B^+ = 0$$

where $\mathbf{Y}^+$ contains the components of the input field inside the triangular cavity.

12

We introduce as well a matrix $\mathbf{S}$ which selects the cavity components $\mathbf{Y}^+$:

$$\begin{bmatrix} 0 \\ 0 \\ \mathbf{Y}^+ \end{bmatrix} = \mathbf{S}\mathbf{X}^+ \quad \Rightarrow \quad S_{ij} = \delta_{ij}(1 - \delta_{11})(1 - \delta_{22})$$

where $\delta_{ij}$ is the Kronecker symbol. We can then express:

$$\mathbf{X}^+ = \mathbf{S}\mathbf{X}^+ + \mathbf{X}^0, \qquad \mathbf{X}^0 = [1, 0, \cdots, 0]^T$$

The two relations between $\mathbf{X}^+$ and $\mathbf{X}^-$ are:

$$\mathbf{X}^- = \mathbf{C}\mathbf{X}^+ \qquad \mathbf{X}^+ = \mathbf{P}\mathbf{X}^-$$

The coupling matrix $\mathbf{C}$ is:

$$\mathbf{C} = \begin{bmatrix} \mathbf{M}_S & & & 0 \\ & \mathbf{M}_C & & \\ & & \ddots & \\ 0 & & & \mathbf{M}_C \end{bmatrix}$$

where there are as many $\mathbf{M}_C$ in the matrix as there are corners in the triangular cavity, not counting the splitter. Note that in the edge-connected cavity, the $(t, \epsilon')$ and $(t', \epsilon'')$ coefficients must be exchanged as compared to the tip-connected cavity. The matrix $\mathbf{P}$ describes the phase relations between the field components inside he cavity.

Finally, we end up solving the following linear system of equations:

$$\mathbf{X}^- = [\mathbf{1} - \mathbf{CSP}]^{-1}\mathbf{C}\mathbf{X}^0$$

from which we extract the transmission and reflection coefficients $T = |D^-|^2$ and $R = |C^-|^2$.

For the tip-coupled cavity, we have the following definitions (see Fig. 6(a)):

$$\mathbf{X}^{\pm} = \begin{bmatrix} A^{\pm} & B^{\pm} & C^{\pm} & D^{\pm} & E^{\pm} & F^{\pm} & G^{\pm} & H^{\pm} \end{bmatrix}^{T}$$

$$\mathbf{C} = \begin{bmatrix} \mathbf{M}_S & & 0 \\ & \mathbf{M}_C & \\ 0 & & \mathbf{M}_C \end{bmatrix} \qquad \mathbf{P} = \begin{bmatrix} 0 & 0 & 0 & 1 & 0 & 0 & 0 & 0 \\ 0 & 0 & 1 & 0 & 0 & 0 & 0 & 0 \\ 0 & 0 & 0 & 0 & 0 & 0 & e^{i\phi_1} & 0 \\ 0 & 0 & 0 & 0 & 0 & e^{i\phi_1} & 0 & 0 \\ 0 & e^{i\phi_1} & 0 & 0 & 0 & 0 & 0 & 0 \\ 0 & 0 & 0 & 0 & 0 & 0 & 0 & e^{i\phi_0} \\ 0 & 0 & 0 & 0 & e^{i\phi_0} & 0 & 0 & 0 \\ e^{i\phi_1} & 0 & 0 & 0 & 0 & 0 & 0 & 0 \end{bmatrix}$$

with $\phi_i = nk_0 L_i$, where $n$ is the effective index of the edge mode and $k_0 = 2\pi F/c$ with $F$ the frequency of the wave and $c$ the speed of light in vacuum.

For the edge-coupled cavity (see Fig. S6(b)):

$$\mathbf{X}^{\pm} = \begin{bmatrix} A^{\pm} & B^{\pm} & C^{\pm} & D^{\pm} & E^{\pm} & F^{\pm} & G^{\pm} & H^{\pm} & I^{\pm} & J^{\pm} \end{bmatrix}^{T}$$

$$\mathbf{C} = \begin{bmatrix} \mathbf{M}'_S & & & 0 \\ & \mathbf{M}_C & & 0 \\ 0 & & \mathbf{M}_C & \\ & 0 & & \mathbf{M}_C \end{bmatrix} \qquad \mathbf{P} = \begin{bmatrix} 0 & 0 & 0 & 1 & 0 & 0 & 0 & 0 & 0 & 0 \\ 0 & 0 & 1 & 0 & 0 & 0 & 0 & 0 & 0 & 0 \\ 0 & 0 & 0 & 0 & 0 & 0 & 0 & 0 & e^{i\phi_2} & 0 \\ 0 & 0 & 0 & 0 & 0 & e^{i\phi_2} & 0 & 0 & 0 & 0 \\ 0 & e^{i\phi_2} & 0 & 0 & 0 & 0 & 0 & 0 & 0 & 0 \\ 0 & 0 & 0 & 0 & 0 & 0 & 0 & e^{i\phi_0} & 0 & 0 \\ 0 & 0 & 0 & 0 & e^{i\phi_0} & 0 & 0 & 0 & 0 & 0 \\ 0 & 0 & 0 & 0 & 0 & 0 & 0 & 0 & 0 & e^{i\phi_0} \\ 0 & 0 & 0 & 0 & 0 & 0 & e^{i\phi_0} & 0 & 0 & 0 \\ e^{i\phi_2} & 0 & 0 & 0 & 0 & 0 & 0 & 0 & 0 & 0 \end{bmatrix}.$$

where $\mathbf{M}'_S$ has the same expression as $\mathbf{M}_S$ but where $\gamma$ and $\delta$ have been exchanged and $\beta$ has been changed into $-\beta$.

# 6 Frequency-dependent coefficients of $\mathbf{M}_C$ and $\mathbf{M}_S$

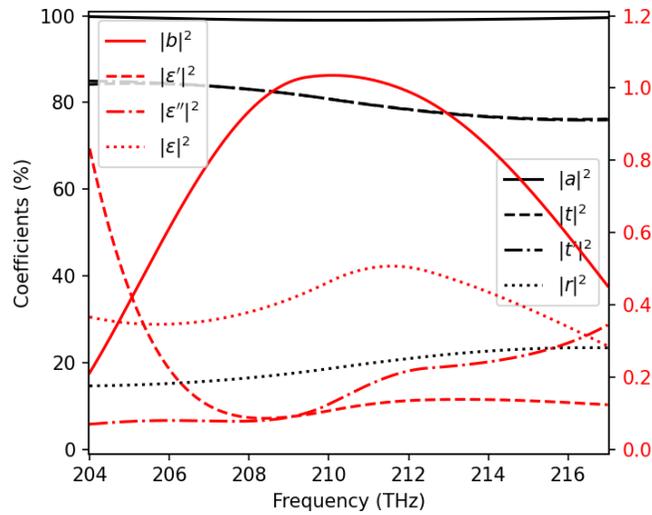

Figure S7: Frequency evolution of the coefficients amplitude associated to $\mathbf{M}_C$ and $\mathbf{M}_S$ after optimization.

15

# 7 Field distribution in a fractal-type resonator

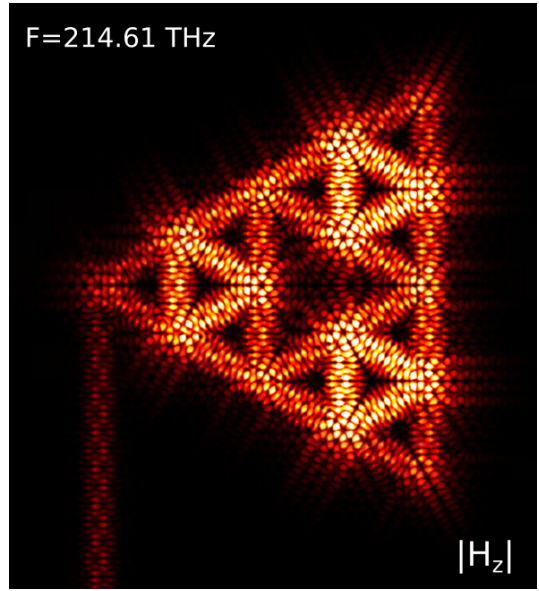

Figure S8: Distribution of the amplitude of the magnetic field at a resonance of the resonator corresponding to the second iteration of the Sierpiński triangle construction.



\end{document}